\documentclass[11pt]{article}

\usepackage{cite}
\usepackage{amssymb,amsmath}
\usepackage{latexsym}
\usepackage{mathrsfs}
\usepackage{hyperref} 
\usepackage{verbatim}
\usepackage{graphicx}

\def\x{{\bf x}}
\def\y{{\bf y}}
\def\z{{\bf z}}
\def\k{{\bf k}}
\def\p{{\bf p}}

\def\R{{\mathcal R}}
\def\P{{\mathcal P}}

\def\bea{\begin{eqnarray}}
\def\eea{\end{eqnarray}}
\def\be{\begin{equation}}
\def\ee{\end{equation}}
\def\ba{\begin{array}}
\def\ea{\end{array}}
\def\nn{\nonumber}

\begin{document}

\setlength\arraycolsep{2pt}

\renewcommand{\theequation}{\arabic{section}.\arabic{equation}}
\setcounter{page}{1}

\begin{titlepage}

\begin{center}

\vskip 1.0 cm

{\LARGE  \bf Untangling features in the primordial spectra}

\vskip 1.0cm

{\large
Gonzalo A. Palma
}

\vskip 0.5cm

{\it
\mbox{Grupo de Cosmolog\'ia y Astrof\'isica Te\'orica}\\
\mbox{Departamento de F\'{i}sica, FCFM, Universidad de Chile}\\ 
\mbox{Blanco Encalada 2008, Santiago, Chile}
}

\vskip 1.5cm

\end{center}

\begin{abstract}

We discuss the possible existence of features in both the primordial power spectrum and bispectrum generated during a stage of single field cosmic inflation. We argue that there are two main classes of features: those produced by a sudden time variation of the sound speed of curvature perturbations, and those produced by a sudden change in the expansion rate during inflation. The former are known to be produced by heavy fields, when the inflationary background trajectory in field space undergoes a bend, whereas the latter are known to be produced by features in the inflaton potential encountered as the inflaton field descends its slope. In general, features are expected to be the result of these two sources combined, however, it is possible that one source dominated over the other, resulting in a distinctive pattern that may be observationally tested. We deduce a relation that gives us the shape of features in the bispectrum provided that we know the shape of features in the power spectrum, and show that each one of these two classes of features leaves a particular footprint in the distribution of perturbations that could be uncovered by a joint analysis of the primordial power spectrum and bispectrum.

\end{abstract}

\end{titlepage}

\newpage

\section{Introduction} 
\setcounter{equation}{0}

Single field slow-roll inflation~\cite{Guth:1980zm, Linde:1981mu, Albrecht:1982wi} predicts a nearly Gaussian distribution of primordial curvature perturbations, characterized by a set of featureless scale-invariant spectra. The measurement of departures from scale-invariance, using Cosmic Microwave Background (CMB) observations, has been key to the understanding of inflation~\cite{Starobinsky:1980te, Mukhanov:1981xt}. It is now clear that the primordial power spectrum is slightly red tilted~\cite{Ade:2013uln, Ade:2015lrj} in a way that is consistent with the fact that the Hubble radius was steadily growing during inflation. In addition, the bispectrum is predicted to be almost scale invariant and suppressed by the slow-roll parameters characterizing inflation~\cite{Maldacena:2002vr, Chen:2006nt}, in perfect agreement with current observations~\cite{Senatore:2009gt, Ade:2013ydc, Ade:2015ava}.  On the other hand, power spectrum reconstructions from current CMB data marginally admit features with a 5\% amplitude modulation, while parametric searches of features allow similar levels of modulations depending on the specific model used to fit data~\cite{Ade:2013uln, Ade:2015lrj}. Although there is no evidence of features in the spectra, their search is both appealing and significant~\cite{Hunt:2013bha, Meerburg:2013cla, Meerburg:2014kna, Hazra:2014jwa, Hu:2014hra, Fergusson:2014hya, Munchmeyer:2014cca, Fergusson:2014tza}. The presence of features in the primordial spectra would definitely rule out the simplest models of inflation, calling for a better understanding of how curvature perturbations were generated and evolved during the epoch of inflation.

Features in the power spectrum are known to appear in a number of circumstances where curvature perturbations interacted with other degrees of freedom during inflation. For instance, in multi-field models of inflation, the inflaton background trajectory has the chance to meander the potential landscape~\cite{Gordon:2000hv, GrootNibbelink:2000vx, GrootNibbelink:2001qt, Tye:2008ef, Tye:2009ff}, enhancing the interaction between curvature perturbations and fields ortogonal to the trajectory each time the trajectory suffered a bend~\cite{Tolley:2009fg, Cremonini:2010sv, Cremonini:2010ua, Achucarro:2010da, Cespedes:2012hu}. The resulting interactions produce localized features in the spectra if the interaction was gentle enough to reduce the sound speed of curvature perturbations without exciting the heavy fields orthogonal to the trajectory~\cite{Achucarro:2010da, Achucarro:2012sm, Cespedes:2012hu, Saito:2013aqa, Gao:2013ota}. In this case, the dynamics of curvature perturbations is well captured by a single effective field theory parametrizing the interaction between curvature perturbations and heavy fields through nontrivial operators, leading to a non-canonical description of inflation~\cite{Achucarro:2010jv, Baumann:2011su, Achucarro:2012sm, Avgoustidis:2012yc, Achucarro:2012yr, Burgess:2012dz, Gwyn:2012mw, Cespedes:2013rda, Castillo:2013sfa, Gwyn:2014doa, Gong:2014rna}. On the other hand, if the bend rate of the trajectory was comparable to ---or larger than--- the mass of the orthogonal heavy fields, these become excited, generating resonant features in the spectra~\cite{Shiu:2011qw, Cespedes:2012hu, Chen:2014cwa}. Other mechanisms to produce features in the spectra include situations where the inflaton field traverses a feature in the potential~\cite{Starobinsky:1992ts, Adams:2001vc, Gong:2005jr, Covi:2006ci, Ashoorioon:2006wc, Ashoorioon:2008qr, Romano:2014kla} (\emph{e.g.} potentials with step features), particle production during inflation~\cite{Chung:1999ve, Elgaroy:2003hp, Mathews:2004vu, Romano:2008rr, Barnaby:2009dd, Fedderke:2014ura}, and scenarios where the inflaton tunneled several local minima of the potential~\cite{Chialva:2008xh} (\emph{e.g.} chain inflation). 

Of course, a mechanism leading to the appearance of features in the power spectrum is bound to generate features in higher order correlation functions~\cite{Chen:2006xjb, Chen:2008wn, Arroja:2011yu, Adshead:2011bw, Adshead:2011jq, Arroja:2012ae}. As a consequence, one expects a correlation between features in the power spectrum and bispectrum~\cite{Achucarro:2012fd, Jackson:2013mka, Adshead:2013zfa, Noumi:2013cfa, Miranda:2013wxa, Gong:2014spa, Ashoorioon:2014yua}. In ref.~\cite{Achucarro:2012fd}, for instance, it was shown that if features in the power spectrum result from the time variation of the sound speed of adiabatic modes (as expected in the case of multi-field models with heavy fields), then it is possible to deduce a map telling us the shape of features in the bispectrum determined by those of the power spectrum. This realization has led to the search of correlated features in the primordial spectra~\cite{Achucarro:2013cva, Adshead:2014sga, Achucarro:2014msa, Hu:2014hra, Fergusson:2014hya, Fergusson:2014tza} by analyzing the latest non-Gaussian constraints coming from CMB observations. On the other hand, it is possible to have background inflationary dynamics where no significant features are generated at the level of two point correlation functions, but with large features at higher order correlation functions~\cite{Khoury:2008wj, Ribeiro:2012ar, Behbahani:2012be}.

The aim of this article is to study the generation of sharp features in both the power spectrum and bispectrum in models of inflation where curvature perturbations are due to a single scalar degree of freedom. We are interested in uncovering well defined relations between the power spectrum and bispectrum, allowing us to asses the origin of features according to their shape and scale dependence. The models we wish to describe include canonical models of inflation with features in the potential, multi-field models with heavy fields orthogonal to the inflaton trajectory~\cite{Achucarro:2010da}, and other non-canonical models of inflation~\cite{ArmendarizPicon:1999rj, Silverstein:2003hf, Alishahiha:2004eh} with nontrivial properties leading to the generation of features. To keep our discussion as general as possible we adopt the Effective Field Theory (EFT) of inflation perspective~\cite{Cheung:2007st, Senatore:2010wk}, wherein one focuses exclusively on the dynamics of fluctuations, providing a systematic parametrization of departures from the standard single field canonical case~\cite{Baumann:2011su, Gwyn:2012mw, Cespedes:2013rda, Gwyn:2014doa}. This approach is ideal to study the generation of features in the spectra, because it attributes the generation of features to the sudden variation of time dependent quantities parametrizing the inflationary background of the theory, for instance, through the time variation of the sound speed~\cite{Takamizu:2013wja, Firouzjahi:2014fda}.

We will show that features in the power spectrum necessarily come together with features in the bispectrum, with a shape determined by the specific dynamical mechanism underlying their appearance. More precisely, we classify the origin of features in the power spectrum as coming from two possible sources: 
\begin{itemize}
\item[(a)] Departures from quasi-de Sitter, that is, by a sudden change in the expansion rate of inflation. These could be due to features in the inflaton potential, leading to sudden changes in the slow-roll parameters. 
\item[(b)] Sudden changes in the sound speed, which in principle may be uncorrelated to the expansion rate. These could appear in multi-field models with heavy fields, when the inflationary background trajectory in field space undergoes a bend.
\end{itemize}
We show that each one of these two sources implies a unique relation between features in the power spectrum and bispectrum, that may be told apart. Of course, more generally we expect a combination of these two sources, leading to more involved relations between the power spectrum and bispectrum. While we will not examine situations where variations of the speed of sound and deviations to quasi de Sitter happen simultaneously in this work, we shall lay the groundwork to study the generation of features in the spectra in more general contexts.

As mentioned, previous works have already studied the correlation of features in the primordial spectra. In ref.~\cite{Achucarro:2012fd}, for instance, a relation was found between the power spectrum and bispectrum for the particular case in which features were due to variations in the sound speed. Moreover, in ref.~\cite{Gong:2014spa} an approach based on the generalized slow-roll method was proposed to examine the correlation of features between different $n$-point correlation functions, which was further analyzed and extended in ref.~\cite{Achucarro:2014msa}. Here, however, we offer an explicit comparison between features generated by different sources, allowing for a direct method to compare data, in the event that sharp features are discovered in future CMB, and large scale structure measurements.

\subsection{Main results} 

Before moving to a more technical and detailed discussion, let us summarize the main results of our work. We will show that if sharp features are present in the power spectrum as the result of a sudden time variation of the background, they will also contribute to the appearance of sharp features in the bispectrum. In particular, if features in the power spectrum are exclusively sourced by a sudden time variation of the sound speed, not affecting the evolution of other slow-roll quantities parametrizing the expansion rate of the inflationary universe, then the equilateral and folded limits of $f_{\rm NL}$, parametrizing the bispectrum, are given by\footnote{In this expressions, $k_1$, $k_2$ and $k_3$ are the magnitude of the momenta on which the bispectrum depends on. The equilateral limit corresponds to the case where $k_1 = k_2 = k_3$, whereas the folded limit corresponds to the case $k_1 = k_2 = k_3/2$ (or its permutations). Finally, the squeezed limit of eq.~(\ref{squeezed}) corresponds to the configuration $k_1 = k_2$ and $k_3 = 0$ (or permutations of it).}
\bea
 f_\textrm{NL-s}^{\rm (eq)} &\simeq&   \frac{5 }{54} \left[ \frac{d^2}{d \ln k^2}  \frac{\Delta \P}{\P_0} (k)  \right]_{k = (k_1 + k_2 + k_3)/2} , \label{sound-equi} \\
 f_\textrm{NL-s}^{\rm (fold)}  &\simeq&   \frac{1}{16} \left[ \frac{d^2}{d \ln k^2}  \frac{\Delta \P}{\P_0} (k)  \right]_{k = (k_1 + k_2 + k_3)/2} , \label{sound-fold}
\eea
where $\Delta \P \equiv \P(k) - \P_0 $ is the difference between the complete power spectrum $\P(k)$ and its featureless counterpart $\P_0$ obtained in the absence of sudden time variations of the background. Similarly, if features in the power spectrum are sourced by a sudden change of the expansion rate during inflation, then the equilateral and folded limits of $f_{\rm NL}$ are found to be:
\bea
 f_{\textrm{NL-}\epsilon}^{\rm (eq)}  &\simeq&   \frac{5 }{36} \left[ \frac{d^2}{d \ln k^2}  \frac{\Delta \P}{\P_0} (k)  \right]_{k = (k_1 + k_2 + k_3)/2} , \label{quasi-equi}  \\ 
 f_{\textrm{NL-}\epsilon}^{\rm (fold)}  &\simeq&  \frac{1}{12} \left[ \frac{d^2}{d \ln k^2}  \frac{\Delta \P}{\P_0} (k)  \right]_{k = (k_1 + k_2 + k_3)/2} . \label{quasi-fold}
\eea
In addition, we find that in both cases the so called squeezed limit of $f_{\rm NL}$ takes the form
\be
 f_{\rm NL}^{\rm (sq)} = - \frac{5}{12}  \left[ \frac{d}{d \ln k}  \frac{\Delta \P}{\P_0} (k)  \right]_{k = (k_1 + k_2 + k_3)/2} , \label{squeezed}
\ee
in agreement with the consistency relation between the power spectrum and the bispectrum~\cite{Maldacena:2002vr, Creminelli:2004yq, Cheung:2007sv}. These two sets of relations represent limits of the generation of features in the power spectra, assuming that one source dominated over the other. More generally, we expect features to be a result of the combined action of the variation of the sound speed and slow roll quantities parametrizing the evolution of the background. 
These results offer a handle to analyze the presence of features in the data. They tell us that the bispectrum depends on the second derivative of the spectrum, multiplied by a coefficient that is unique to the source of the feature. This coefficient is larger for those features generated as the result of a sudden time variation of the expansion rate of the inflationary universe.

\subsection{Outline} 

This article is organized as follows: In Section~\ref{sec:prim-pert} we review the dynamics of primordial perturbations originated during inflation, and define the power spectrum and bispectrum. In Section~\ref{sec:pert-th} we review the in-in formalism of perturbation theory necessary to compute the primordial spectra taking into account the effects induced by sudden time variations of the background. Then, in Section~\ref{sec:features} we deduce the analytical expression determining the shape of features in the power spectrum due to sudden changes of the background. We also show how to invert this relation to obtain an analytical expression giving the time dependence of the background in terms of the features in the power spectrum. Then, in Section~\ref{sec:bispectrum} we use the results of the previous sections to derive an analytical expression relating features in the power spectrum with features in the bispectrum. Finally, in Section~\ref{sec:conclusions} we discuss our main results and offer some concluding remarks. In addition, in Appendix~\ref{app:formulae} we summarize some useful formulas allowing for the computation of the bispectrum as a function of the features in the power spectrum.


\section{Primordial perturbations}  \label{sec:prim-pert}
\setcounter{equation}{0}

Let us start by setting the formalism through which we will study the evolution of perturbations and compute features. We will take the effective field theory of inflation point of view, by which one focuses exclusively on the perturbations and disregard the dynamics of the background~\cite{Cheung:2007st}. The advantage of adopting this perspective is that any information concerning the background is parametrized by time dependent quantities in the action describing perturbations, therefore freeing us from worrying about the specific model that led to such time dependence. First, we adopt the co-moving gauge and use the Arnowitt-Deser-Misner (ADM) formalism~\cite{Arnowitt:1962hi} to write the metric as
\be
ds^2 = - N^2 dt^2 + a^2(t) e^{ 2 \R } \delta_{i j} (d x^i + N^i dt)(d x^j + N^j dt) ,
\ee
where $a(t)$ is the scale factor, parametrizing the expansion of inflation. In the previous expression, $\R$ represents the adiabatic spatial curvature perturbation. In addition, $N$ and $N^i$ are the lapse and shift functions respectively. After solving the constraint equations for the lapse and shift functions in terms of $\R$, and inserting them back into the action for adiabatic curvature perturbations, one deduces an action of the form $S = S_{2} + S_{3}$, where the quadratic part is given by
\be
S_{2} = m_{\rm Pl}^2 \int \! d^4 x \,  a^3 \epsilon \left[ \frac{1}{c_s^2} \dot {\mathcal{R}}^2 - \frac{1}{a^2}(\nabla \mathcal{R})^2  \right] , \label{R-2-action}
\ee
where $\epsilon = - \dot H / H^2$ is the first slow roll parameter, with $H = \dot a / a$ the Hubble expansion rate. On the other hand, $c_s$ is identified with the sound speed at which perturbations propagate at long wavelengths. Furthermore, the cubic contribution to the action is found to be~\cite{Seery:2005wm}
\bea
S_{3} &=& m_{\rm Pl}^2 \int \! d^4 x \,  a^3 \epsilon \bigg[ 
\frac{3}{c_s^2} \R \dot \R^2 - \frac{1}{a^2} \R (\nabla \R)^2  - \frac{1}{H c_s^2}  \left(1 +  \frac{ 2 \lambda c_s^2 }{ \epsilon H^2} \right)  \dot \R^3  \nn \\
&& +  \frac{1}{2 \epsilon a^4} \left\{ \left( 3 \R - \frac{1}{H}\dot \R \right) \left[ \partial^i \partial^j \psi \, \partial_i \partial_j \psi  - (\nabla^2 \psi)^2 \right] - 4  \nabla^2 \psi \,  \partial^i \R \, \partial_i \psi \right\}
 \bigg] , \label{S-R-3}
\eea
where $\psi$ is the scalar part of the shift function defined through $N_i = \partial_i \psi$, and given in terms of $\R$ by
\be
\nabla^2 \psi =  \frac{a^2 \epsilon}{c_s^2} \dot \R - \frac{1}{H} \nabla^2 \R . \label{psi-constraint}
\ee
In~(\ref{S-R-3}), the quantity $\lambda$ has mass-squared dimensions and parametrizes physics beyond the canonical picture of single-field slow-roll inflation. In order to study the generation of features in the bispectrum, it is useful to rewrite the action~(\ref{S-R-3}) by performing certain integrations by parts~\cite{Burrage:2011hd}. One obtains:
\bea
S_{3} &=& m_{\rm Pl}^2 \int \! d^4 x \,  a^3 \epsilon \bigg[ 
\frac{1}{c_s^4} \left[ 3 (c_s^2 - 1) + \epsilon - \eta \right] \R \dot \R^2 + \frac{1}{c_s^2 a^2} \left( (1 - c_s^2) + \eta + \epsilon - 2  s  \right) \R (\nabla \R)^2  \nn \\
&&  + \frac{1}{H }  \left(\frac{1 - c_s^2}{c_s^4}  - \frac{ 2 \lambda }{ \epsilon H^2} \right)  \dot \R^3  + \frac{1}{4 a^4} (\partial \chi)^2 \nabla^2 \R - \frac{4 - \epsilon}{2 \epsilon a^4} \nabla^2 \chi \partial^i  \R \partial_i \chi + \frac{f }{\epsilon a^3} \frac{\delta S_2}{\delta \R}
 \bigg] , \label{S-R-3-second}
\eea
where $\chi$ is defined as $\chi \equiv \psi + \R / H$ and, because of (\ref{psi-constraint}), it satisfies $\nabla^{2} \chi = a^2 \epsilon \dot \R /  c_s^{2}$. In addition, we have defined $s \equiv \dot c_s / H c_s$. Notice the appearance of $\eta$, which is the standard second slow-roll parameter hereby defined as:
\be
\eta = \frac{\dot \epsilon}{H \epsilon}.
\ee
On the other hand, the function $f$, multiplying the linear equation of motion $\delta S_2 / \delta \R$ is found to be given by:
\bea
f &=& - \frac{1}{4 a^2 H} \bigg[ 	\frac{4 a^2}{c_s^2} \dot \R^2 - (\partial \R)^2 +  \partial^i \chi \partial_i \R \nn \\
&& + \nabla^{-2} \partial^i \partial^j \left( \partial_i \R \partial_j \R \right) - \nabla^{-2} \partial^i \left( \nabla^2 \R \partial_i \chi + \nabla^2 \chi \partial_i \R \right) \bigg] .
\eea
The advantage of working with the cubic action of eq.~(\ref{S-R-3-second}) instead of (\ref{S-R-3}) is that $\eta$ appears explicitly as a coupling. We will exploit this fact to deduce how departures from quasi-de Sitter generate features in the bispectrum.

\subsection{Canonically normalized perturbations}

In order to compute $n$-point correlation functions it is convenient to work with conformal time $\tau$, defined through the change of coordinates $d t = a d \tau$, and introduce a canonically normalized field $u(\x , \tau)$ through the following field-redefinition:
\be
u = z  \, \mathcal{R}, \qquad z = \sqrt{2 \epsilon} \frac{a}{c_s}. \label{u-R}
\ee
Inserting this redefinition back into eq.~(\ref{R-2-action}), and choosing units such that $m_{\rm Pl}^2 = 1$, we arrive to the following quadratic action for the $u$-field
\be
S_{2} =  \frac{1}{2} \int d^3x \, d \tau  \left[ (u')^2 - c_s^2 (\nabla u)^2 + \frac{z''}{z} u^2 \right] , \label{S-u}
\ee
where the prime denotes derivatives with respect to conformal time $'= \partial_\tau$, and $z'' /z$ is found to be given by:
\bea
\frac{z''}{z} &=&  a^2 H^2 \left( 2 - \epsilon + \frac{\eta}{2} - s \right) \left( 1 + \frac{ \eta}{2} - s \right) + a  H \left( \frac{\eta'}{2} - s' \right) . \label{z-prime-prime}
\eea 
The $u$-field corresponds to the canonically normalized adiabatic curvature perturbation, and its canonical momentum is given by $\pi = \partial \mathcal L / \partial u'$, which to leading order reads $\pi = u'$. Both $u$ and $\pi$ are required to satisfy the equal time commutation relation
\be
\left[ u(\x,\tau) , \pi (\y, \tau) \right] = i \delta(\x - \y) . \label{quant-commut-rel}
\ee
where $\delta(\x - \y)$ is the usual Dirac delta function. We shall solve this system perturbatively in Section~\ref{sec:pert-th} where we define our strategy to split the theory into free and interacting parts, appropriate for the study of features in the spectra.  

\subsection{Power spectrum}

In order to define the power spectrum, and later on the bispectrum, let us write the adiabatic perturbations in Fourier space by introducing the mode function $\hat \R_{\k} (\tau) $ as:
\be
\R (\x,\tau) = \frac{1}{(2 \pi)^{3}} \int d^3 k \, \hat \R_{\k} (\tau) e^{i \k \cdot \x} .
\ee
Then, the power spectrum $P (k,\tau)$, evaluated at a given time $\tau$, is defined via the equation:
\be
\langle \hat \R_{\k} (\tau) \hat \R_{\p} (\tau) \rangle = (2 \pi )^3 \delta (\k + \p) P (k , \tau) . \label{power-def}
\ee
It is common practice to work with the dimensionless power spectrum $\mathcal P_{\mathcal R} ( k , \tau)$ instead of $P(k,\tau)$ which is defined as:
\be
\mathcal P_{\mathcal R} ( k , \tau) = \frac{k^3}{2 \pi^2 } P(k,\tau) .
\ee
Putting together (\ref{u-R}) and (\ref{power-def}) we deduce an expression for the power spectrum $\mathcal P_{\mathcal R} ( k , \tau)$ in terms of the two-point correlation function of the $u$-field
\be
\mathcal P_{\mathcal R} ( k , \tau) =  \frac{k^3}{2 \pi^2  z^2}   \int_x \langle u(\x,\tau) u(0,\tau)  \rangle  e^{- i \k \cdot  \x} ,\label{power-spectrum-def}
\ee
where $\int_x$ stands for $\int d^3 x$. Notice that one of the fields has been conveniently evaluated at $\y = 0$, allowed by the homogeneity and isotropy of the background. The computation of $\langle u(\x,\tau) u(0,\tau)  \rangle$ will depend on how we decide to split the action into free and interaction parts. We will come back to this question in Section~\ref{sec:pert-th} where we advocate for a particular splitting that will be particularly helpful to analyze features.

\subsection{Bispectrum}

Let us now move on to consider the three point correlation function. The bispectrum $B_{\R }(\k_1 , \k_2 , \k_3, \tau)$ is conventionally defined as:
\be
 \langle \hat \R_{\k_1} (\tau) \hat \R_{\k_2} (\tau) \hat \R_{\k_3} (\tau) \rangle = (2 \pi)^3 \delta (\k_1 + \k_2 + \k_3 ) B_\R (\k_1 , \k_2 , \k_3, \tau) .
\ee
This expression may be inverted to give the bispectrum in terms of the three-point correlation function of the $u$-field as
\be
B_\R (\k_1 , \k_2 ,\k_3 , \tau) = \frac{1}{z^3}  \int_x  \int_y  \langle u (\x,\tau) u (\y,\tau) u (0,\tau) \rangle e^{- i \k_1 \cdot \x - i \k_2 \cdot \y} , \label{bispectrum-def}
\ee
where $\k_3 = - \k_2 - \k_1$. Just as we did with eq.~(\ref{power-spectrum-def}), we have evaluated one of the fields at the comoving coordinate $\z = 0$. We will come back to these expressions during the next sections.

\subsection{The long wavelength limit}

To finish this discussion, let us recall that our goal is to compute correlation functions in the long wavelength limit $| \tau k | \ll 1$, which gives us the spectra at the end of inflation. Thus, the power spectrum and bispectrum we are interested in correspond to the following formal limits:
\bea
\P_{\R} ( k) &\equiv& \lim_{|\tau| \to 0} \P_{\R} ( k , \tau) , \\
B_\R (\k_1 , \k_2 ,\k_3 ) &\equiv& \lim_{|\tau| \to 0} B_\R (\k_1 , \k_2 ,\k_3 , \tau) .
\eea
Given that $\R$ is constant after horizon crossing, these expressions give us the initial conditions for adiabatic perturbations outside the horizon for the hot Big-Bang era, in terms of background parameters at horizon crossing time, during inflation.


\section{Perturbation theory} \label{sec:pert-th}
\setcounter{equation}{0}

We now discuss the computation of the power spectrum and bispectrum using perturbation theory as a tool. Our aim is to provide a systematic procedure allowing us to compute the effects of features in both the power spectrum and bispectrum, incorporating effects of sudden deviations from quasi-de Sitter order by order. We will achieve this by using the in-in formalism in a quasi-de Sitter space-time~\cite{Maldacena:2002vr, Weinberg:2005vy}. 

\subsection{Splitting the theory}

To begin with, we define de Sitter space-time as the zeroth order geometry parametrizing the background. We do this by writing the zeroth order scale factor as
\be
a_0(\tau) = - \frac{1}{H_0 \tau} ,
\ee
where $H_0$ is the zeroth order Hubble expansion rate of the de Sitter space-time (assuming that we are slicing de Sitter with the help of cosmological coordinates). This allows us to write
\be
 \frac{z''}{z} \equiv  \frac{z_0''}{z_0} \left[ 1 + \frac{1}{2} \delta (t)  \right], \qquad \frac{z_0''}{z_0} = \frac{2}{\tau^2} . \label{delta-def}
\ee
where $z_0''/z_0$ is the zeroth order contribution to $z''/z$, and $\delta (t)$ parametrizes its deformations. We may proceed similarly with the sound speed, and define
\be
\theta (\tau)  \equiv 1 - c_s^2 (\tau) , \label{theta-def}
\ee
as the time dependent function parametrizing the sudden variation of the sound speed $c_s$.
Notice that $\theta(\tau) \geq 0$, in order to avoid unphysical situations where $c_s > 1$. In addition, in the absence of evidence of a sound speed different from unity,\footnote{A suppressed sound speed $c_s$ is known to produce a sizable level of non-Gaussianity~\cite{Chen:2006nt, Senatore:2009gt, Ade:2013ydc}, so far undetected. In addition, it is known that a large value of the tensor to scalar ratio $r$ disfavors small values of the sound speed~\cite{Baumann:2014cja, Palma:2014faa}. } we will keep our analysis within the regime of small departures $\theta \ll 1$. Both $\delta$ and $\theta$ parametrize deviations from the background values $z_0'' / z_0 = 2 / \tau^2$ and $c_s = 1$, however, it is important to notice that since $z$ depends on $c_s,$ then $\delta$ already contains a contribution that depends on variations of $\theta$. Indeed, if one retains linear terms with respect to $\theta$, one finds\footnote{In order to write eq.~(\ref{complete-delta}) we have only kept linear contributions in $\theta$ to keep our discussion simple. The more accurate expression has terms quadratic in $\theta$ that \emph{in principle} could become large in later computations (\emph{c.f.} Section~\ref{sec:features-power-spectrum}). However, we have verified that indeed such terms remain subleading.}
\be
\delta = \delta_{H} -  \tau \theta'  + \frac{\tau^2}{2} \theta'' ,Ê\label{complete-delta}
\ee
where $\delta_{H}$ represents deviations from quasi-de Sitter which are independent of $c_s$.  We will come back to the particular form of $\delta_H$ in a moment. Having defined the zeroth order background, we may now split the full action $S = S_2 + S_3$ into two sectors as
\be
S = S_{0} + S_{\rm int} ,
\ee
with the zeroth order part $S_0$ given by:
\be
S_0 = \frac{1}{2} \int d^3x \, d \tau  \left[ (u')^2 - (\nabla u)^2 + \frac{z_0''}{z_0} u^2 \right].
\ee
This splitting implies that the interaction part $S_{\rm int}$ of the action contains both quadratic and cubic contributions
\be
S_{\rm int} = S_{\rm int}^{(2)}  + S_{\rm int}^{(3)} , \label{interaction-action}
\ee
where the quadratic interaction part is given by:
\be
S_{\rm int}^{(2)} =  \frac{1}{2}\int d^3x \, d \tau  \left[   \theta ( \tau )  (\nabla u)^2 + \frac{1}{\tau^2}   \delta (\tau)  u^2 \right].
\ee
On the other hand, the cubic contribution is given by eq.~(\ref{S-R-3-second}) with the appropriate replacements of 
$\R$ written in terms of $u$, dictated by eq.~(\ref{u-R}). Equation~(\ref{interaction-action}) implies a quadratic contribution to the interaction part of the Hamiltonian, which is found to be given by:
\be
H_{\rm int}^{(2)} (\tau) = - \frac{1}{2} \int d^3x  \left[   \theta ( \tau )  (\nabla u)^2 + \frac{\delta (\tau) }{\tau^2} u^2 \right] . \label{Hamiltonian-2-int}
\ee
There is also a cubic contribution of the form $H_{\rm int}^{(3)} (\tau)$ which is easily obtained by taking into account $S_{\rm int}^{(3)}$ in eq.~(\ref{interaction-action}). We shall come back to this term in Section~\ref{sec:features}.

\subsection{The in-in formalism}

Perturbation theory tells us that the field $u(x,\tau)$ and its canonical momentum $\pi (x,\tau)$, satisfying the commutation relation~(\ref{quant-commut-rel}), may be written in terms of the interaction-picture field $u_I(x,\tau)$ evolved in time with the help of the propagator $U(\tau)$ as:
\bea
&& u(\x,\tau) = U^{\dag}(\tau) \,  u_I(\x,\tau) \, U(\tau) , \label{u-complete} \\
&& \pi (\x,\tau) = U^{\dag}(\tau) \,  u_I' (\x,\tau) \, U(\tau) . \label{pi-complete}
\eea
Here the interaction-picture field $u_I(x,\tau)$ has the form of a free field, which may be expanded in Fourier modes as
\be
u_I (\x,\tau) = \frac{1}{(2 \pi)^{3}} \int d^3 k \left[ a_\k u_k(\tau) e^{i \k \cdot \x}  + a_\k^{\dag} u_k^{*}(\tau) e^{-i \k \cdot \x}   \right] , \label{u-I-pict}
\ee
where $a_\k^{\dag}$ and $a_\k$ are creation and annihilation operators satisfying the usual commutation relations for particle states in a Fock space:
\be
\big[ a_\k , a_\p^{\dag} \big] = (2\pi)^3 \delta(\k - \p). 
\ee
On the other hand, $u_k(\tau)$ represents the normalized solution to the zeroth order equation of motion in momentum space:
\be
u_{k}'' + \left( k^2  - \frac{2}{\tau^2} \right) u_{k} = 0. \label{eq-of-motion}
\ee
The solution to this equation is obtained by selecting the Bunch-Davis vacuum, and is given by:
\be
u_k(\tau) = \frac{1}{\sqrt{2 k}} \left( 1 - \frac{i}{k \tau} \right) e^{- i k \tau} . \label{BD-solution}
\ee
Returning to~(\ref{u-complete}), the propagator is given by
\be
U(\tau) = \mathcal T \exp \left\{ - i \int^{\tau}_{-\infty_+} \!\!\!\!\!\! d\tau' H_I (\tau') \right\} ,
\ee
where $ \mathcal T$ stands for the standard time ordering symbol and $\infty_+ = (1 + i \epsilon) \infty$ is the prescription isolating the in-vacuum in the infinite past. Furthermore, $H_I$ is the interaction Hamiltonian derived from~(\ref{interaction-action}) expressed in terms of the interaction-picture field as:
\be
H_I (\tau) = U^\dag(\tau)  H_{\rm int}  U(\tau) . \label{def-int-pict-ham}
\ee
For instance, putting together eq.~(\ref{Hamiltonian-2-int}) and (\ref{def-int-pict-ham}) we immediately find that the quadratic contribution to the interaction-picture Hamiltonian has the form:
\be
H_{I}^{(2)} (\tau) = - \frac{1}{2} \int d^3x  \left[   \theta ( \tau )  (\nabla u_I)^2 + \frac{\delta (\tau) }{\tau^2} u_I^2 \right] . \label{int-pict-ham-2}
\ee
It is useful to express this result in momentum space. We do this by defining $\hat u_{\k} (\tau)$ as the Fourier transform of the $u_I$-field of eq.~(\ref{u-I-pict}) in the following way:
\be
u_I (\x,\tau) = \frac{1}{(2 \pi)^{3}} \int_k\, \hat u_{\k} (\tau) e^{i \k \cdot \x} , \qquad \hat u_{\k} (\tau) \equiv a_\k u_k(\tau)  + a_{-\k}^{\dag} u_k^{*}(\tau),
\ee
where $\int_k$ stands for $\int d^3 k$. Then, the Hamiltonian of eq.~(\ref{int-pict-ham-2}) becomes
\be
H_{I}^{(2)} (\tau) =  \frac{1}{(2 \pi)^{3}} \int_k  k^2  h_{k}(\tau)  \hat u_{\k} (\tau)  \hat u_{-\k} (\tau)  , \label{Hamilton-quad}
\ee
where $h_{k}(\tau)$ is a dimensionless function parametrizing the interaction, given by:
\be
h_{k}(\tau) \equiv -  \frac{1}{2} \left[  \theta ( \tau )    + \frac{1}{ k^2 \tau^2} \delta ( \tau ) \right] . \label{h_k-def}
\ee
One may work out the form of $H_I^{(3)}$ in momentum space in a similar way.

\subsection{$n$-point correlation functions}

The present formalism allows us to compute $n$-point correlation functions to any desired order in terms of the parameters $\delta$ and $\theta$. To be more precise, an $n$-point correlation function for the $u$-field, at a fixed time $\tau$, is given by:
\be
\langle u(\x, \tau) \cdots u(\y, \tau)  \rangle = \langle 0 | U^{\dag}(\tau) \,  u_I(\x,\tau) \cdots u_I(\y,\tau) \, U(\tau) | 0 \rangle. \\
\ee
Then, by expanding the propagator $U(\tau)$ to second order in $H_I$, we obtain:
\bea
\langle u(\x, \tau) \cdots u(\y, \tau)  \rangle &=&  \langle 0| u_I(\x,\tau) \cdots u_I(\y,\tau) | 0 \rangle  - 2 \Im \int^{\tau}_{-\infty_-} \!\!\!\!\!\! d\tau'   \langle 0|  H_I(\tau') u_I(\x,\tau) \cdots u_I(\y,\tau) | 0 \rangle \nn \\
&& - 2 \Re \int^{\tau}_{-\infty_-} \!\!\!\!\!\! d\tau'  \int^{\tau'}_{-\infty_-} \!\!\!\!\!\! d\tau''  \langle 0|  H_I(\tau'') H_I(\tau') \, u_I(\x,\tau) \cdots u_I(\y,\tau) | 0 \rangle \nn \\ 
&& + \int^{\tau}_{-\infty_-} \!\!\!\!\!\! d\tau' \int^{\tau}_{-\infty_+} \!\!\!\!\!\! d\tau''   \, \langle 0| H_I(\tau') \,  u_I(\x,\tau) \cdots u_I(\y,\tau) \, H_I(\tau'') | 0 \rangle .
\eea
In the present work, we are only interested in retaining first order contributions in terms of $\delta$ and $\theta$, to the power spectrum and bispectrum. To compute the power spectrum, we need the two-point correlation function, which is found to be given by
\be
\langle u(\x, \tau) u(\y, \tau)  \rangle =  \langle 0| u_I(\x,\tau) u_I(\y,\tau) | 0 \rangle + i \int^{\tau}_{-\infty} \!\!\!\!\!\! d\tau' \,  \langle 0|  \left[ H_I^{(2)}(\tau') ,u_I(\x,\tau) u_I(\y,\tau) \right] | 0 \rangle,   \label{two-point-pert}
\ee
where $\langle 0| u_I(\x,\tau) u_I(\y,\tau) | 0 \rangle$ corresponds to the zeroth order two-point correlation function for $u$. Putting together (\ref{power-spectrum-def}), (\ref{u-I-pict}) and (\ref{BD-solution}), it is found to be given by:
\be
\langle 0| u_I( \x , \tau) u_I( \y , \tau) | 0 \rangle  = \frac{1}{(2 \pi)^3} \int d^3 k |u_k(\tau)|^2 e^{i \k \cdot ( \x - \y)} . 
\ee
From this result, one obtains the zeroth order power spectrum of adiabatic curvature perturbations:
\bea
\P_0 (k) \equiv \lim_{\tau \to 0} \frac{k^3}{2 \pi^2 z^2} | u_k(\tau) |^2  = \frac{H_0^2 }{8  \pi^2 \epsilon_0}  .  \label{zeroth-power}
\eea
On the other hand, to compute the bispectrum, we need the three-point correlation function, which is found to be given by:
\be
\langle u(\x, \tau) u(\y, \tau)  u(\z, \tau) \rangle =   i \int^{\tau}_{-\infty} \!\!\!\!\!\! d\tau'  \, \langle 0|  \left[ H_I^{(3)} (\tau') , u_I(\x,\tau) u_I(\y,\tau) u_I(\z,\tau)  \right] | 0 \rangle.  \label{three-point-pert}
\ee
Equations~(\ref{two-point-pert}) and~(\ref{three-point-pert}) constitute the basis of the computations that will take place in the following sections.


\section{Features in the power spectrum} \label{sec:features}
\setcounter{equation}{0}

In this section we compute how a sudden variation of the background affects the shape of the primordial power spectrum. Let us start by specifying more precisely what we mean by a sudden variation of the background. 

\subsection{Sudden variations of the background} \label{sec:sudden-back}

First, we shall disregard slow-roll corrections to the background and simply focus on first order contributions due to variations of $\delta_H ( \tau )$ and $\theta ( \tau )$ larger than those implied by standard slow-roll. Having clarified this, by a sudden variation of the background we specifically refer to  variations of both $\delta_H ( \tau )$ and $\theta ( \tau )$ satisfying
\be
\left|  \frac{d \delta_H }{dN} \right| \gg \left| \delta_H \right|, \qquad  \left|  \frac{d \theta }{dN} \right| \gg \theta , \label{sudden-def}
\ee
where $N \equiv \ln a$ measures $e$-folds. By keeping $|\delta_H ( \tau )| \ll 1$ and $|\theta ( \tau )| \ll 1$ at all times, eq.~(\ref{sudden-def}) tells us that $\delta_H$ and $\theta$ changed over a characteristic period $\Delta N$ much smaller than an $e$-fold. Notice that another way to express these relations is
\be
\left| \tau \delta_H' \right| \gg \left| \delta_H \right|, \qquad  \left| \tau \theta' \right| \gg \theta . \label{delta-H-theta-N}
\ee
Given that we are assuming that slow roll inflation is not interrupted, at least in average, we will assume that deviations from quasi-de Sitter (parametrized by $\delta_H$) are such that $\epsilon$ has the form:
\be
\epsilon = \epsilon_0 + \Delta \epsilon (\tau), \qquad  \epsilon \ll 1,
\ee
where $\epsilon_0$ is a constant for all practical purposes. Then, the evolution of the background may be such that sudden variations of $\Delta \epsilon (\tau)$ induce variations of $\eta$ that satisfy
\be
|\eta| \ll 1, 
\ee
but with large sudden variations of the form:
\be
\left| \tau  \eta'  \right| \gg \left| \eta \right| .
\ee
Then, by recalling eqs.~(\ref{z-prime-prime}) and~(\ref{delta-def}), we see that under the present assumptions $\delta_H$ is dominated by a variation of $\eta$ in the following way:
\be
\delta_H = -  \frac{1}{2} \tau \eta' . \label{delta-H-eta}
\ee
Having established this point, we will further assume that sudden variations happen during a finite period of time, in such a way that $h_k(\tau)$, defined in eq~(\ref{h_k-def}), satisfies
\be
\lim_{k |\tau| \to 0} h_{k} (\tau) \to 0 , \qquad  \lim_{k |\tau| \to  \infty} h_k (\tau) \to 0 ,
\ee
sufficiently fast. 

\subsection{Features in the power spectrum due to sudden variations} \label{sec:features-power-spectrum}

Let us now compute the features in the power spectrum due to sudden variations of the background. Our starting point is eq.~(\ref{two-point-pert}) which expresses the two-point correlation function to first order in $h_k (\tau)$ defined in eq.~(\ref{h_k-def}). Taking the Fourier transform of eq.~(\ref{two-point-pert}) we obtain the following expression for the power spectrum
\be
\P_{\mathcal R} (k)  = \P_{0} + \Delta \P(k) ,
\ee
where $\mathcal P_{0}$ represents the zeroth order contribution found in (\ref{zeroth-power}), and $\Delta \mathcal P(k)$ embodies deviations due to $h_k (\tau)$, given by
\bea
\frac{\Delta \mathcal P}{\mathcal P_0 } (k) = \lim_{k |\tau| \to 0}   i ( 2 c_0^3 k^3 \tau^2 )  \int^{\tau}_{-\infty} \!\!\!\!\!\! d\tau' \, \int_y  e^{- i \k \cdot  \y}   \langle 0|  \left[ H_I(\tau') ,u_I(\y,\tau) u_I(0,\tau) \right] | 0 \rangle .   \label{comp-1}
\eea
Putting together eqs.~(\ref{Hamilton-quad}) and~(\ref{h_k-def}) back into the spatial integral of eq~(\ref{comp-1}), allows us to find:
\bea
 && \int_y  e^{- i \k \cdot  \y}  \langle 0|  \left[ H_I(\tau') ,u_I(\y,\tau) u_I(0,\tau) \right] | 0 \rangle = \nn \\
&& \qquad 2  i   \frac{h_k (\tau')}{k^2} \Bigg[  \left( \frac{1}{ k \tau} - \frac{1}{ k \tau'} \right)\left( 1 + \frac{1}{ k^2 \tau \tau'} \right)  \cos (2  k ( \tau' - \tau))  \nn \\ 
&& \qquad + \frac{1}{2} \left\{   \left( \frac{1}{ k \tau} - \frac{1}{ k \tau'} \right)^2  -\left( 1 + \frac{1}{ k^2 \tau \tau'} \right)^2  \right\}  \sin (2  k ( \tau' - \tau))  \Bigg] .
\eea
Then, performing the change of variables $x = - k \tau$, we deduce
\bea
 \frac{\Delta \mathcal P}{\mathcal P_0 } (k) &=& \lim_{x \to 0}     4 x^2   \int^{\infty}_{x} \!\!\!\!\!\! dx' \,  h_k (-x'/k) \Bigg[   \left( \frac{1}{ x} - \frac{1}{ x'} \right) \left( 1 + \frac{1}{ x x'} \right)  \cos (2  (x' - x))  \nn \\ 
&&  + \frac{1}{2} \left\{   \left( \frac{1}{ x} - \frac{1}{ x'} \right)^2  -\left( 1 + \frac{1}{ x x'} \right)^2  \right\}  \sin (2   ( x' - x))  \Bigg]   . \label{comp-2}
\eea
Since $h_k (\tau)$ is assumed to vanish quickly in the limits $|k \tau| \to 0$ and $|k \tau| \to  \infty$, we may perform the limit $x \to 0$ in eq.~(\ref{comp-2}) before performing  integral.\footnote{If $h_k (\tau)$ does not vanish in the limit $|k \tau| \to 0$ one is forced to solve the integral before computing the limit. This happens, for instance, in computations where $h_k (\tau)$ is dominated by slow-roll parameters, here assumed to be negligible.} This step gives (after dropping the prime $'$):
\be
 \frac{\Delta \mathcal P}{\mathcal P_0 } (k) =    4  \int^{\infty}_{0} \!\!\!\!\!\! dx \,  h_k (-x/k) \Bigg[   \frac{1}{ x}   \cos (2  x )  + \frac{1}{2} \left[   1 -  \frac{1}{ x^2}  \right]  \sin (2  x)  \Bigg]   .
\ee
Performing a partial integration to the term containing the cosine, the previous expression may be reduced to
\be
 \frac{\Delta \mathcal P}{\mathcal P_0 } (k) =  2  \int^{\infty}_{0} \!\!\!\!\!\! dx \, \left[  h_k (-x/k) -  \frac{1}{x} \frac{\partial }{\partial x} h_k (- x/k) \right]  \sin (2  x) . \label{Delta-P-x}
\ee
To complete the computation, we recall that $\delta(\tau)$ contains a contribution due to the variation of the sound speed represented by $\theta$ [\emph{c.f.} eq.~(\ref{complete-delta})] explicitly given by:
\be
\delta = \delta_{H} - \tau \theta'  + \frac{\tau^2}{2} \theta''  .
\ee
Using this expression in eq.~(\ref{h_k-def}), one finds that $h_k (- x /k)$ has the form:
\be
h_k (- x /k) =  - \frac{1}{2} \left[  \theta  - \frac{1 }{x}  \frac{\partial }{\partial x} \theta + \frac{1}{2}  \frac{\partial^2 }{\partial x^2} \theta  + \frac{1}{x^2} \delta_{H} \right].
\ee
Inserting this expression back into eq.~(\ref{Delta-P-x}), and performing several partial integrations, we finally obtain:
\bea
\frac{\Delta \mathcal P}{\mathcal P_0 } (k) &=&  k \int_{-\infty}^{0} \!\!\! d \tau \, \left[  - \theta + \frac{\delta_H}{k^2Ê\tau^2}  + \frac{2 \delta_H}{k^4Ê\tau^4}  - \frac{1}{k^4Ê\tau^3} \frac{d \delta_H}{d \tau} \right] \, \sin (2 k \tau) . \label{features-delta}
\eea
This represents one of our main results. Equation (\ref{features-delta}) gives us the shape of features in the power spectrum as a function of the sudden variations of the background parametrized by $\delta_H$ and $\theta$.

\subsection{Inverting the power spectrum}

Result (\ref{features-delta}) may be inverted to obtain the functional forms of $\delta_H$ and $\theta$ necessary to have features $\Delta \mathcal P$ of certain $k$-dependence. First, we formally extend $\delta_H (\tau)$ and $\theta (\tau)$ from the domain $-\infty < \tau \leq 0$ to the domain $-\infty < \tau < + \infty$, by demanding that both $\delta_H (\tau)$ and $\theta (\tau)$ are odd functions under the exchange $\tau \to - \tau$.\footnote{This idea was introduced in ref.~\cite{Achucarro:2012fd} to invert the power spectrum as a function of the sound speed using a Fourier transformation.} This extension allows us to rewrite (\ref{features-delta}) as
\bea
\frac{\Delta \mathcal P}{\mathcal P_0 } (k) &=&  \frac{k}{2} \int_{-\infty}^{+\infty} \!\!\! d \tau \, \left[  - \theta + \frac{\delta_H}{k^2Ê\tau^2}  + \frac{2 \delta_H}{k^4Ê\tau^4}  - \frac{1}{k^4Ê\tau^3} \frac{d \delta_H}{d \tau} \right] \, \sin (2 k \tau) . \label{features-delta-2}
\eea
Next, we multiply eq.~(\ref{features-delta-2}) by $k^3$ and perform several partial integrations to trade powers of $k$ for time derivatives inside the integral of the right hand side. This gives us:
\be
k^3 \frac{\Delta \mathcal P}{\mathcal P_0 } (k) = - \frac{1}{4} \int_{-\infty}^{+\infty} \!\!\! d \tau \, \left[ \frac{1}{8} \theta'''' + \frac{ \delta_H'' }{ 2Ê\tau^2}  - \frac{ \delta_H}{ \tau^4}   \right] \, \sin (2 k \tau) \, .
\ee
This expression may now be inverted to give
\bea
\frac{1}{8} \theta'''' +  \frac{ \delta_H'' }{ 2Ê\tau^2}  - \frac{ \delta_H}{ \tau^4}  = \frac{4 }{\pi i}  \int_{-\infty}^{+\infty} \!\!\! d k \, k^3 \frac{\Delta \mathcal P}{\mathcal P_0 } (k)  e^{- 2 i k \tau} . \label{inverted-power}
\eea
This result informs us about the time dependence of $\delta_H$ and $\theta$ needed to produce features $\Delta \mathcal P$ of a given shape. As expected, there is a degeneracy between $\delta_H$ and $\theta$ that precludes us from deducing where do features come from based on measurements of the power spectrum alone. To gain some insight on eq.~(\ref{inverted-power}) let us study the cases in which features are produced by variations of the sound speed and the expansion rate separately.

\subsection{Features due to a sudden change in the sound speed}

Let us first examine the case in which features are exclusively due to a sudden deviation of the sound speed from its value $c_s = 1$. Setting $\delta_H = 0$ in eq.~(\ref{inverted-power}), and reducing the number of time derivatives in the left hand side, by decreasing the powers of $k$ in the right hand side, we obtain:
\bea
 \theta (\tau) = \frac{2 }{\pi i}  \int  \frac{d k}{k}  \frac{\Delta \mathcal P}{\mathcal P_0 } (k)  e^{- 2 i k \tau} . \label{theta-P}
\eea
This expression coincides with the one deduced in ref.~\cite{Achucarro:2012fd} where a different method was used to achieve it. Equation (\ref{theta-P}) allows us to infer the time dependence of $c_s(\tau)$ in order to obtain features of known shape $\Delta \mathcal P (k)$.

\subsection{Features due to a sudden change of the expansion rate}

Next, we examine the case in which features are exclusively due to a sudden time variation of the background without a contribution coming from the sound speed. By setting $\theta = 0$ in eq.~(\ref{inverted-power}) we first obtain:
\be
\frac{ \delta_H'' }{ 2Ê\tau^2}  - \frac{ \delta_H}{ \tau^4}  = \frac{4}{i \pi }  \int \! d k \, k^3 \frac{\Delta \mathcal P}{\mathcal P_0 } (k)  e^{- 2 i k \tau} . \label{delta-H-P}
\ee
To exploit this expression, let us recall that we are interested in sudden features in the spectra originated by variations of the background satisfying eq.~(\ref{delta-H-theta-N}). Such variations imply that the second term of the left hand side of~(\ref{delta-H-P}) is subdominant, and therefore may be disregarded:
\bea
\frac{ \delta_H'' }{ 2Ê\tau^2}   = \frac{4}{i \pi }  \int \! d k \, k^3 \frac{\Delta \mathcal P}{\mathcal P_0 } (k)  e^{- 2 i k \tau} .
\eea
Moreover, using~(\ref{delta-H-eta}) and neglecting subdominant terms again, we finally arrive to
\be
\eta(\tau) =  \frac{1}{i \pi }  \int \! d k  \, \frac{d}{dk} \frac{\Delta \mathcal P}{\mathcal P_0 } (k)    e^{- 2 i k \tau} ,
\ee
which gives us the map between features in the power spectrum and $\eta$. This is the desired result linking sudden deviations from quasi-de Sitter with features in the power spectrum.

\subsection{The shape of features induced by sudden background variations} \label{sec:scale-of-feature}

To finish this section, notice that the conditions $\left| \tau \theta' \right| \gg \theta$ and $\left| \tau \eta' \right| \gg \eta$ discussed in Section~\ref{sec:sudden-back} are both equivalent to
\be
\left| \frac{d}{d \ln k} \frac{\Delta \P}{\P_0}  \right| \gg \left| \frac{\Delta \P}{\P_0}  \right| . \label{sharp-features-hierarchy}
\ee
This implies that a sudden variation of the background induces features $\Delta \P / \P_0$ with a characteristic scale $\Delta k$ of oscillation such that $\Delta k \ll k$. This result will turn out to be relevant in our analysis of features in the bispectrum.


\section{Features in the bispectrum}  \label{sec:bispectrum}
\setcounter{equation}{0}

We now examine the effects of sudden variations of $\theta(\tau)$ and $\delta_H (\tau)$ on the bispectrum. Our aim is to deduce an expression for the bispcetrum $B(\k_1, \k_2, \k_3)$ of the form
\be
B (\k_1, \k_2, \k_3) = B_0 (\k_1, \k_2, \k_3) + \Delta B (\k_1, \k_2, \k_3) ,
\ee
where $B_0$ represents the featureless zeroth order expression suppressed by slow roll parameters~\cite{Chen:2006nt}, and $\Delta B$ represents the features resulting from the sudden variation of $\theta(\tau)$ and $\delta_H (\tau)$. Let us begin by identifying the leading contributions to the third order action of eq.~(\ref{S-R-3-second}). In the first place, we notice that the coefficient in front of $\dot \R^3$ in the second line of eq.~(\ref{S-R-3-second}) is naturally a parameter of order:
\be
\frac{1-c_s^2}{c_s^4} - \frac{2 \lambda}{\epsilon H^2} \sim  \frac{(1 - c_s^2)^2}{c_s^4}.
\ee
To appreciate this more clearly, we notice that such a term is proportional to the expansion parameter $M_3^4$ used in the effective field theory approach introduced by Cheung {et al.}~\cite{Cheung:2007st} to parametrize single field inflation. It was further argued in ref.~\cite{Senatore:2009gt} that this term is naturally of order $(1 - c_s^2)^2/c_s^4$, which is found to be the case in explicit examples such as DBI inflation~\cite{Alishahiha:2004eh} and single field EFT's of multi-field models with heavy fields~\cite{Achucarro:2012sm, Cespedes:2013rda}. All of this implies that the term proportional to $\dot \R^3$ in eq.~(\ref{S-R-3-second}) contributes terms quadratic in $\theta$ and therefore may be disregarded.  In second place, let us observe that the term proportional to $f$ in eq.~(\ref{S-R-3-second}) may be removed by a field redefinition, which finally contributes a term in $\Delta B$ which is proportional to $f$ evaluated at time $\tau$. However, since we are assuming that both $\theta(\tau)$ and $\delta_H (\tau)$ vanish at $\tau = 0$, this term does not contribute to the final result. Last but not least, let us recall that, as we have already argued in Section~\ref{sec:sudden-back}, the main contribution to the variation of the expansion rate is determined by $\eta$. Putting all of these considerations together, eq.~(\ref{S-R-3-second}) reduces to:
\be
S_{3} = - \int \! d^4 x \,  a^3 \epsilon  \bigg\{   (3 \theta + \eta) \R \dot \R^2 - \frac{1}{a^2} ( \theta - \tau \theta' + \eta)  \R (\nabla \R)^2   \bigg\} . 
\ee
This form of the cubic action allows us to deduce the cubic contribution to the cubic Hamiltonian in the interaction picture $H_{\rm I}^{(3)} $, which is found to be
\be
H_{I}^{(3)} =  \int_x   a^3 \epsilon \bigg\{   (3 \theta + \eta)  \R_I \dot \R_I^2 - \frac{1}{a^2} ( \theta - \tau \theta' + \eta)   \R_I (\nabla \R_I)^2   \bigg\} , \label{H-int-definite}
\ee
where $\R_I = u_I / z_0$. Before finishing this discussion, notice that we have kept the term $\theta$ in the coefficient in front of the operator $\R_I (\nabla \R_I)^2$, which, according to our discussion of Section~\ref{sec:sudden-back}, is subleading in comparison to $- \tau \theta'$. We have opted to keep this term in order to be able to compare our results with those of ref.~\cite{Achucarro:2012fd}, where features due to sudden variations in the sound speed were studied, but no assumptions on the sharpness of features were made. In addition, one might wonder whether $\eta$ should be taken to be suppressed with respect to $- \tau \theta' $. Here we will make no assumptions about the relative rate of variation of $\eta$ and $\theta$, and wait to see the contribution of each term in the final result to be deduced in the next subsections.

Now that we have the interaction Hamiltonian, we may proceed to compute the induced features in the bispectrum by using the expressions of Appendix~\ref{app:formulae}. We will proceed by examining each source separately.

\subsection{Features due to a sudden change in the sound speed}

Let us start by analyzing the appearance of features due to sudden changes in the sound speed. Putting together eq.~(\ref{H-int-definite}) with the equations of the Appendix~\ref{app:formulae} for the particular case $\eta = 0$, we obtain the desired bispectrum with the form $\Delta B = B_2 + B_3$. The first contribution $B_2$ may be obtained by using eq.~(\ref{B-5}) with $\alpha_2 = 3 \theta$, and expressing $\theta$ in terms of the power spectrum with the help of eq.~(\ref{theta-P})
\bea
B_2  &=& \frac{ ( 2 \pi)^4 \P_0^2}{ (k_1 k_2 k_3)^{2}}  \bigg[ \frac{3}{8} \frac{k_2 k_3 + k_3 k_1 + k_1 k_2}{k^2}\left( \frac{d}{d \ln k}   \frac{\Delta \mathcal P}{\mathcal P_0 } (k)  -  \frac{\Delta \mathcal P}{\mathcal P_0 } (k)  \right)  \nn \\  
&& - \frac{3}{4} \frac{(k_2 k_3)^2 + (k_3 k_1)^2 +(k_1 k_2)^2}{k_1 k_2 k_3} \frac{1}{k}  \frac{\Delta \mathcal P}{\mathcal P_0 } (k)    \bigg]_{k = K/2}  ,
\eea
where $K= k_1 + k_2 + k_3$. The second contribution $B_3$ is obtained similarly by employing eq.~(\ref{B-6}) with $\alpha_3 = \theta - \tau \theta' $:
\bea
B_3  &=&  \frac{ (2 \pi)^4 \P_0^2}{ (k_1 k_2 k_3)^{3}}   \frac{k_1^2 + k_2^2 + k_3^2}{8} \bigg[   \frac{1}{2} \left( \frac{1}{k^2}  \frac{d^2}{d \ln k^2} \frac{\Delta \mathcal P}{\mathcal P_0 } (k) - \frac{1}{k^2} \frac{\Delta \mathcal P}{\mathcal P_0 } (k) \right)   k_1 k_2 k_3 \nn \\Ê
&& -   \left( \frac{1}{k} \frac{\Delta \mathcal P}{\mathcal P_0 } (k) + \frac{1}{k} \frac{d}{d\ln k} \frac{\Delta \mathcal P}{\mathcal P_0 } (k)   \right)  ( k_1 k_2 + k_2 k_3+ k_3 k_1) + 4 k \frac{\Delta \mathcal P}{\mathcal P_0 } (k)  \bigg]_{k = K/2}  .
\eea
The final expression  $\Delta B = B_2 + B_3$ coincides with that obtained in ref.~\cite{Achucarro:2012fd}, and may be simplified significantly by recalling that sharp features in the power spectrum satisfy eq.~(\ref{sharp-features-hierarchy}). This implies that the bispectrum is dominated by the highest derivatives with respect to $\ln k$, leading to the final expression
\bea
\Delta B &=& \frac{ ( 2 \pi)^4 \P_0^2}{ 16 (k_1 k_2 k_3)^{2}} \frac{k_1^2 + k_2^2 + k_3^2}{k^2} \bigg[ \frac{d^2}{d \ln k^2} \frac{\Delta \mathcal P}{\mathcal P_0 } (k) \nn \\
&&
\qquad\qquad - 2 k    \frac{k_1 k_2 + k_2 k_3+ k_3 k_1}{k_1 k_2 k_3}    \frac{d}{d\ln k} \frac{\Delta \mathcal P}{\mathcal P_0 } (k)  \bigg]_{k = K/2} , \label{B-sound}
\eea
where we have kept a term containing the first order derivative which become dominant in the squeezed limit $k_3 \ll k_1 , k_2$. 

\subsection{Features due to a sudden variation of the expansion rate}

We now consider the case of features due to a sudden variation of the expansion rate. Here we must use the results of Appendix~\ref{app:formulae} with eq.~(\ref{H-int-definite}) after setting $\theta = 0$, to finally obtain a bispectrum of the form $\Delta B =  B_2 + B_3$. The first contribution $B_2$ is obtained by using eq.~(\ref{B-5}) with $\alpha_2 = \eta$ in companion of eq.~(\ref{theta-P})
\bea
B_2 &=& \frac{(2\pi)^4 \P_0^2}{ (k_1 k_2 k_3)^{2}} \bigg[ \frac{ k_2 k_3 + k_1 k_2 + k_3 k_1 }{16 \, k^2}  \frac{d^2}{d \ln k^2} \frac{\Delta \mathcal P}{\mathcal P_0 } (k)  \nn \\
&& -  \frac{(k_2 k_3)^2 + (k_3 k_1)^2 +(k_1 k_2)^2}{8 \, k_1 k_2 k_3  \, k }  \frac{d}{d \ln k} \frac{\Delta \mathcal P }{\mathcal P_0 } (k) \bigg]_{k=K/2}  ,
\eea
where $K= k_1 + k_2 + k_3$. On the other hand, the second contribution $B_3$ is obtained by using eq.~(\ref{B-6}) with $\alpha_3 = \eta$:
\be
B_3 =  \frac{ (2 \pi)^4 \P_0^2}{ (k_1 k_2 k_3)^{3}} (k_1^2 + k_2^2 + k_3^2 )\bigg[  \frac{ k_1 k_2 k_3}{32 \, k^2}  \frac{d^2}{d \ln k^2}  \frac{\Delta \mathcal P}{\mathcal P_0 } (k)  - \frac{ k_1 k_2 + k_2 k_3+ k_3 k_1}{16 \, k} \frac{d}{d \ln k} \frac{\Delta \mathcal P}{\mathcal P_0 } (k)  \bigg]_{k=K/2} .
\ee
Notice that to write both of these results we have only kept second derivatives of $\Delta \P / P_0$ with respect to $\ln k$ (which dominate the expression) and first derivatives that become important in the squeezed limit. Adding up these two results, we then obtain:
\bea
\Delta B &=& \frac{(2\pi)^4 \P_0^2}{ 8 (k_1 k_2 k_3)^{3}} \bigg[  (k_1 k_2 k_3)   \frac{d^2}{d \ln k^2} \frac{\Delta \mathcal P}{\mathcal P_0 } (k) \nn \\
&&  -  \bigg( 2 ( k_1^2 + k_2^2 + k_3^2) k - k_1 k_2 k_3  - k_1^3 - k_2^3 - k_3^3 \bigg)  \frac{d}{d \ln k} \frac{\Delta \mathcal P}{\mathcal P_0 } (k)   \bigg]_{k=K/2} . \label{B-de-sitter}
\eea
This expression may be now compared with that of eq.~(\ref{B-sound}) to distinguish features coming from a sudden variation of the sound speed v/s a sudden variation of the quasi-de Sitter background. We compare and discuss both results in the next section.


\section{Discussion \& conclusions}   \label{sec:conclusions}
\setcounter{equation}{0}

Although the analysis of current CMB data offers no evidence of features in the primordial spectra, a better grasp of how they could be uncovered is central to a robust understanding of cosmic inflation. The presence of sharp features in the primordial spectra would rule out the simplest models of inflation, and provide us with precious information about the nature of the fundamental theory embedding inflation. If present, they would strongly support the notion that curvature perturbations interacted with other degrees of freedom during inflation, favoring multi-field models of inflation~\cite{Achucarro:2010da, Shiu:2011qw, Cespedes:2012hu, Saito:2013aqa, Gao:2013ota, Battefeld:2013xka, Battefeld:2013xwa, Battefeld:2014aea}, and leave the door open to consider even more exotic classes of models~\cite{Mizuno:2014jja}. They would also offer a formidable challenge for fundamental theories such as supergravity and string theory, which are highly constrained by the gravitational nature of their couplings~\cite{Kachru:2003sx, Burgess:2004kv, Conlon:2005jm, Silverstein:2008sg, Flauger:2008ad, Baumann:2011nk, McAllister:2008hb, Covi:2008ea, Covi:2008cn, Hardeman:2010fh, Kallosh:2010xz, Borghese:2012yu, Roest:2013aoa}, and require a better understanding of the ultraviolet nature of inflation~\cite{Chialva:2014paa}.

In this work, we have analyzed the relation between features in the power spectrum and bispectrum in models of inflation characterized by sudden time-variations of the inflationary background. We have tried to keep our discussion as general as possible, taking as our main assumption that curvature perturbations are due to a single scalar degree of freedom. This allowed us to adopt the effective field theory approach to parametrize inflation. From a technical point of view, our main results are summarized by eqs.~(\ref{B-sound}) and~(\ref{B-de-sitter}), showing the dependence of the bispectrum on second derivatives of the power spectrum in the two limits where the features are sourced either by a sudden change in the sound speed or by a sudden change in the expansion rate.  To have some clarity about the size of the features in the bispectrum, it is useful to introduce the standard $f_{\rm NL}$ parameter by writing:
\be
 f_{\rm NL} \equiv \frac{10}{3} \frac{k_1 k_2 k_3 }{k_1^3 + k_2^3 + k_3^3} \frac{(k_1 k_2 k_3)^2 }{(2 \pi)^4 \P_\R^2} \Delta B  .
\ee
Since we are interested in first order effects coming from features, in this expression $\P_\R$ stands for the featureless power spectrum $\P_0$. This leads to the results already highlighted in the introduction, namely, that $ f_{\rm NL}$ is proportional to the second derivative of the power spectrum with respect to $\ln k$:
\be
 f_{\rm NL}  \simeq  \beta  \left[ \frac{d^2}{d \ln k^2}  \frac{\Delta \mathcal P}{\mathcal P_0} (k)  \right]_{k = (k_1 + k_2 + k_3)/2} , \label{fnl-beta}
\ee
where $\beta = \beta (k_1 , k_2, k_3 )$ is a smooth function, independent of the overall scale, but dependent on the shape of interest and the source of the feature. To be more precise, if the features are sourced by a sudden variation of the sound speed, we find that $\beta $ is given by:
\be
\beta_\textrm{s} = \frac{5}{6} \frac{k_1 k_2 k_3}{k_1^3 + k_2^3 + k_3^3}    \frac{ k_1^2 + k_2^2 + k_3^2 }{(k_1 + k_2 + k_3)^2}   .\label{beta-s}
\ee
On the other hand, if the features are sourced by a sudden variation of the expansion rate, we find that $\beta $ is given by:
\be
\beta_{\epsilon} =  \frac{5}{12} \frac{k_1 k_2 k_3}{k_1^3 + k_2^3 + k_3^3}   . \label{beta-epsilon}
\ee
The function $\beta$ is an observable, and may be constrained by tests of features in the non-Gaussian angular spectrum~\cite{Ade:2013ydc, Fergusson:2014hya, Fergusson:2014tza}. One may also assess the value of $f_{\rm NL}$ in specific directions of the wavenumber space. For instance, in the case of features generated by sound speed variations, we recover the results of eqs.~(\ref{sound-equi}) and~(\ref{sound-fold}) for the equilateral and folded limits respectively, which in terms of $\beta$ may be expressed as:
\be
\beta_\textrm{s}^{\rm (eq)}  =  \frac{5 }{54} \simeq 0.09, \qquad \qquad \beta_\textrm{s}^{\rm (fold)}  =  \frac{1}{16} \simeq 0.06 ,
\ee
whereas if the features are sourced by a sudden variation of the expansion rate, the same limits give the results of eqs.~(\ref{quasi-equi}) and~(\ref{quasi-fold}), which in terms of $\beta$ are given by:
\be
\beta_\epsilon^{\rm (eq)}  =  \frac{5 }{36} \simeq 0.14 , \qquad \qquad \beta_\epsilon^{\rm (fold)}  =  \frac{1}{12} \simeq 0.08 .
\ee
On the other hand, in the squeezed limit one recovers the standard consistency relation~(\ref{squeezed}) between the two spectra, independently of the source generating the feature~\cite{Sreenath:2014nca}. It may be seen that the numerical coefficient $\beta$ is always larger for the case in which the features are sourced by a sudden change in the expansion rate, especially in the equilateral limit, where it attains its maximum value.

Equations (\ref{beta-s}) and (\ref{beta-epsilon}) constitute our main results, and we wish to emphasize the importance of the fact that they are markedly different. As already stressed, we have opted to analyze the generation of features in the spectra due to both, sound speed and expansion rate variations separately, in order to arrive to simple and well defined expressions such as (\ref{beta-s}) and (\ref{beta-epsilon}) without free parameters to adjust. Nevertheless, we envisage that eq.~(\ref{fnl-beta}) represents a good ansatz characterizing the most general situation where features are produced by a mixed effect coming from both types of sources. In that case, $\beta (k_1 , k_2, k_3 )$ will inevitably introduce at least one adjustable parameter connecting both limits (\ref{beta-s}) and (\ref{beta-epsilon}), as it will be shown in a future work~\cite{Palma_features_2}.


\subsection*{Acknowledgements}

I would like to thank Ana Ach\'ucarro, Vicente Atal, Jinn-Ouk Gong, Bin Hu, Sander Mooij, Grigoris Panotopoulos, Subodh Patil, Alex Soto and Jes\'us Torrado for useful discussions and comments on the content of this work. I'm particularly grateful to Pablo Ortiz for giving me valuable insights after carefully reading an earluer version of this manuscript.
This work was supported by the Fondecyt project number 1130777 and by the ``Anillo'' project ACT1122 funded by the ``Programa de Investigaci\'on Asociativa", Conicyt.


\begin{appendix}

\renewcommand{\theequation}{\Alph{section}.\arabic{equation}}

\section{Useful formulae} \label{app:formulae}
\setcounter{equation}{0}

In this appendix we summarize the formulae used to compute the bispectra of Section~\ref{sec:bispectrum}. Let us consider an interaction Hamiltonian of the form:
\be
H_{I} = - \int_x \,  a^3 \epsilon  \bigg\{ \alpha_1 \frac{1}{H_0}  \dot \R_I^3  -  \alpha_2 \, \R_I \dot \R_I^2 + \alpha_3  \frac{1}{a^2}   \R_I (\nabla \R_I)^2   \bigg\} ,
\ee
where $\alpha_1$, $\alpha_2$ and $\alpha_3$ are functions of time, and $\R_I$ is the curvature perturbation in the interaction picture, which has the explicit form
\be
\R_I = \frac{1}{(2 \pi)^3} \int_k \hat\R _\k (\tau)  e^{i \k \cdot \x} ,
\ee
with:
\be
\hat\R _\k (\tau) \equiv a_\k \R_k(\tau)  + a_{-\k}^{\dag}  \R_k^{*}(\tau) , \qquad  \R_k (\tau) = i \frac{H_0}{2 \sqrt{\epsilon_0 k^3 }} \left(1 +  i k \tau \right) e^{- i k \tau} . \label{R-I}
\ee
Then, using eqs.~(\ref{bispectrum-def}) and ~(\ref{three-point-pert}) together, the bispectrum $B (\k_1, \k_2, \k_3)$ is found to have the form
\be
B = B_1 + B_2 + B_3 ,
\ee
with each contribution given by
\bea
B_1(\k_1, \k_2, \k_3) \! &=& \!  \frac{6 \epsilon_0}{ i H_0^2} \R_1(0) \R_2 (0) \R_3(0)  \! \int_{- \infty}^{0} \!\!\!\!\!\! d \tau \frac{\alpha_1(\tau)}{\tau}   \left[ \R_1' (\tau) \R_2' (\tau) \R_3' (\tau) \right]^*  + {\rm c.c.} ,  \label{B-1} \\
B_2(\k_1, \k_2, \k_3) \! &=& \!   \frac{2 \epsilon_0}{i H_0^2} \R_1(0) \R_2 (0) \R_3(0)  \! \int_{- \infty}^{0} \!\!\!\!\!\! d \tau \frac{\alpha_2 (\tau)}{\tau^2} \left[ \R_1 (\tau) \R_2' (\tau) \R_3' (\tau) + sym \right]^* + {\rm c.c.} , \label{B-2} \\
B_3(\k_1, \k_2, \k_3) \! &=& \! \frac{2 \epsilon_0}{i H_0^2} \R_1(0) \R_2 (0) \R_3(0)  \! \int_{- \infty}^{0} \!\!\!\!\!\! d \tau \frac{\alpha_3 (\tau)}{\tau^2}   \left[ \k_2 \cdot \k_3 \,  \R_1 (\tau) \R_2 (\tau) \R_3 (\tau) + sym \right]^* + {\rm c.c.} ,  \nn \\ \label{B-3}
\eea
where $\R_i (\tau) \equiv \R (\k_i, \tau)$. Another useful relation allowing the computation of $B_3$ is
\be
\k_1 \cdot \k_2 + \k_2 \cdot \k_3 +\k_3 \cdot \k_1  =  - \frac{1}{2} ( k_1^2 + k_2^2 + k_3^2) ,
\ee
which is valid as long as $\k_1 + \k_2 + \k_3 = 0$. To obtain more explicit expressions for $B_1$, $B_2$ and $B_3$ we may use eq.~(\ref{R-I}) and the fact that the functions $\alpha_1$, $\alpha_2$ and $\alpha_3$ are odd functions of conformal time $\tau$. Then eqs.~(\ref{B-1}), (\ref{B-2}) and~(\ref{B-3}) become
\bea
B_1  \! &=& \!  -  \frac{ 6 \pi^4 \P_0^2 }{ (k_1 k_2 k_3)}  \! \int_{- \infty}^{+\infty} \!\!\!\!\!\! d \tau  \tau^2  \alpha_1(\tau)  i e^{ i K \tau}  ,  \label{B-4}  \\
B_2   \! &=& \!    \frac{2 \pi^4 \P_0^2 }{(k_1 k_2 k_3)^{2}}  \! \int_{- \infty}^{+\infty} \!\!\!\!\!\! d \tau \alpha_2 (\tau) \left[  i \tau  (k_1 k_2+ k_3 k_1+ k_2 k_3) -  \frac{k_1^2 k_2^2+ k_3^2 k_1^2+ k_2^2 k_3^2}{k_1 k_2 k_3}  \right] i e^{ i K \tau} , \label{B-5} \\
B_3 \! &=& \!   \frac{\pi^4 \P_0^2 }{ (k_1 k_2 k_3)^{3}}  ( k_1^2 + k_2^2 + k_3^2)   \! \int_{- \infty}^{+\infty} \!\!\!\!\!\! d \tau \frac{\alpha_3 (\tau)}{\tau^2}   \left(1 - i k_1 \tau \right)  \left(1 -  i k_2 \tau \right)  \left(1 -  i k_3 \tau \right) i e^{ i K \tau} , \qquad  \label{B-6}
\eea
where $K = k_1 + k_2 + k_3$, and where we have identified  $\P_0 = H_0^2 / 8  \pi^2 \epsilon_0 $.

\end{appendix}

\end{document}